# Thermal Rectification of Solid-liquid Phase Change Thermal Diode under the Effect of Supercooling

Zhaonan Meng, Raza Gulfam, Peng Zhang*, Fei Ma

Institute of Refrigeration and Cryogenics, Shanghai Jiao Tong University, No. 800 Dongchuan Road, Shanghai 200240, China

*Author to whom correspondence should be addressed: zhangp@sjtu.edu.cn

**Abstract**

Thermal rectification ratios of solid-liquid phase change thermal diodes (SL-PCTDs) are not sustainable beyond a certain temperature bias, necessitating reasonably compatible designs. Manipulating the heat transfer in the forward and reverse directions of the SL-PCTD is of great practical interest to achieve persistent performance. Herein, a SL-PCTD prototype is investigated under the supercooling effect of a phase change material (calcium chloride hexahydrate). In the forward direction, supercooling effect plays a significant role in sustaining natural convection far below the melting temperature (30 °C), which leads to robust heat transfer within a temperature bias of 10–40 °C. While in the reverse direction, dislodging the supercooling via manual supercooling release (MSR) within a large temperature range of 30–7 °C tends to greatly inhibit heat transfer. As a consequence, thermal rectification ratio of 2.95 is achieved, which is competitively sustainable at large temperature bias compared to the SL-PCTD without supercooling effect. To address the effect of supercooling on thermal rectification comprehensively, thermal resistance approach is applied to theoretically model the SL-PCTD, which shows a good agreement with experimental data. Such a device with supercooling control in only one direction is further named as the SL-PCTD with gating functionality, where manipulating supercooling effect of phase change material opens up a feasible avenue for enabling generalized thermal diode.

Thermal diode is a bi-terminal nonlinear thermal element which enables heat flow in one direction at a forward temperature bias while blocking heat flow at a reverse temperature bias. This unidirectional and preferential behavior of heat flow is called thermal rectification, which is analogous to an electronic diode, and can be employed for efficient thermal computing and information processing,[1,2] thermal energy harvesting,[3] thermal logic circuitry,[4] and so on. The ratio of a steady-state forward heat flux ($q_{for}$) to a steady-state reverse heat flux ($q_{rev}$) at the same temperature bias is defined as thermal rectification ratio ($\gamma$).[5] In general, the direction with larger heat flux is regarded as the forward direction, and the direction with smaller heat flux being the reverse direction. Therefore, $\gamma \geq 1$ always holds true.

To rectify the heat flow through a thermal diode, variation of the thermal resistance across the thermal medium is essential. This functionality can be achieved either by one material (thermal medium) consisting of asymmetrical structure[6-13] or by two materials[14-34] having different thermo-physical properties. For the latter case, phase change materials (PCMs) provide a robust variation of thermal conductivity during phase change, and such thermal diodes are referred to as phase change thermal diodes (PCTDs). It is imperative to understand various multiphase heat transfer modes of PCMs. For example, solid-solid[14-21] phase change operates by heat conduction or radiation; liquid-vapor[22-23] phase change is driven by thermal convection; and solid-liquid[24-29] phase change is driven by the combination of heat conduction and natural convection. The physical mechanisms of solid-solid and liquid-vapor PCTDs are well understood, leading to widespread applications including thermal management of electronic devices,[18] solar energy utilization[23,24] and refrigeration system[25]. By contrast, the research on solid-liquid PCTDs (SL-PCTDs) is in infant stage, and the dominant physical mechanism is still elusive because of the multiple heat transfer modes, leading to difficulties in thermal media selection and device design. Only until recently was a SL-PCTD composed of calcium chloride hexahydrate ($CaCl_2.6H_2O$) and paraffin wax reported to have achieved a thermal rectification ratio as high as 3.0 (arguably the highest in the family of the SL-PCTDs).[32] However, this thermal rectification ratio cannot be sustained at large temperature biases. Briefly, not only the highest value of thermal rectification ratio is essential, but also the persistent thermal rectification ratio over a wide temperature bias range is of much more practical interest.

Natural convection in the thermal media of a SL-PCTD is the dominant heat transfer mode, which is however diminished as soon as solid phase appears with regard to the operational temperature. The earlier emergence of solid phase is thus a critical issue, which might result in a narrow temperature range for sustainable performance. However, a few PCMs are capable of suppressing the emergence of solid phase far below the conventional melting temperature,[35,36] providing supercooling effect. Previously, natural convection in thermal diodes has been theoretically modeled without supercooling effect,[5,32] leaving the unexplored potential technological advances through exploiting supercooling effect in the SL-PCTDs.

In this letter, a SL-PCTD under the effect of supercooling is theoretically modeled and experimentally tested. To assemble two terminals of the SL-PCTD, calcium chloride hexahydrate and paraffin are utilized. Calcium chloride hexahydrate ($CaCl_2 \cdot 6H_2O$) shows supercooling with a melting temperature of 30 °C and a freezing temperature of 7 °C (see Section 1 in supplementary material), while paraffin shows no supercooling with a melting temperature of 35 °C. Possible variations of the effective thermal conductivity resulted from the involvement of natural convection are accurately described by introducing freezing temperature in the theoretical modeling. Utilizing manual supercooling release (MSR), the thermal rectification ratio of the SL-PCTD is gated effectively in a specified temperature bias range, leading to a maximum thermal rectification ratio of 2.95.

Based on the theoretical analysis of the previous work,[32] a SL-PCTD with an optimum length ratio of 0.12 comprising terminal A (paraffin) and terminal B ($CaCl_2 \cdot 6H_2O$) is fabricated (see Figure S2(a) and S2(b) in supplementary material), and a custom-built experimental setup[32] is used to measure the steady-state heat flux of the SL-PCTD (see Figure S2(c) in supplementary material) in the forward direction (top terminal A and bottom terminal B) and the reverse direction (top terminal B and bottom terminal A), respectively. Under either condition, the temperature ($T_h$) of the bottom hot end of the SL-PCTD is kept constant at 40 °C, while the temperature ($T_c$) of the top cold end of the SL-PCTD is decreased from 30 °C to 0 °C, inducing the supercooling effect of $CaCl_2 \cdot 6H_2O$ to emerge naturally during the cooling process in the experiments. Note that the supercooling effect was absent in the previous work[32] because $T_c$ was increased from 0 °C to 30 °C. Correspondingly, the temperature bias is defined as $\Delta T = T_h - T_c$ under steady-state operation. To inhibit the heat flux of the SL-PCTD in the reverse direction, MSR is adopted by seeding solid crystals of $CaCl_2 \cdot 6H_2O$ into terminal B (see Figure S2(d) in supplementary material) in the $T_c$ range of 30–7 °C, converting supercooled liquid PCM B into solid phase.[37]

To theoretically depict the performance of the SL-PCTD, unidirectional thermal resistance models were built with[5,32] and without[38] considering natural convection in liquid PCM. In the present study, the theoretical models are modified further to take the supercooling effect into account. The heat flux $q$ ($q_{for}$ and $q_{rev}$) through the SL-PCTD is assumed unidirectional, which can be depicted by the thermal resistances in series with uniform heat flux:

$$q = \frac{\Delta T_{As}}{R_{As}} = \frac{\Delta T_{Al}}{R_{Al}} = \frac{\Delta T_{Bs}}{R_{Bs}} = \frac{\Delta T_{Bl\_eff}}{R_{Bl\_eff}} \quad (1)$$

where $R_{As}$, $R_{Al}$, $R_{Bs}$ and $R_{Bl\_eff}$ are the thermal resistances of solid PCM A, liquid PCM A, solid PCM B and liquid PCM B, respectively, and $\Delta T_{As}$, $\Delta T_{Al}$, $\Delta T_{Bs}$ and $\Delta T_{Bl\_eff}$ are the corresponding temperature differences.

Following the operating modes of the SL-PCTD, the theoretical model can be divided into multiple thermal modules having different $q$ ($q_{for}$ and $q_{rev}$) and $\gamma$. In the thermal modules that systematically address the heat transfer of two phases, $T_h$ should be higher than the phase change temperatures ($T_{mA}$ and $T_{mB}$) of PCM A and PCM B, while $T_c$ should be lower than $T_{mA}$ and $T_{mB}$. The thermal modules of the SL-PCTD in the forward and reverse directions have been presented in the previous work,[32] where the supercooling effect was absent. Supercooling keeps on impeding freezing of PCM B until the freezing temperature ($T_{fB}$) is reached. The thermal modules of the SL-PCTD can be classified by the states of PCM A and PCM B which change with $T_c$. Based on the interfacial temperature ($T_i$) between terminals A and B, the states of two PCMs can be determined by comparing $T_i$ with $T_c$, $T_{mA}$, $T_{mB}$ and $T_{fB}$. If supercooling effect is allowed in the reverse direction, PCM B (near the top cold end of the SL-PCTD) retains the (supercooled) liquid phase, while PCM A (near the bottom hot end of the SL-PCTD) exists in liquid phase or solid-liquid two-phase. Consequently, two more additional thermal modules (Reverse S1 and S2) are created as compared with the conventional designs without supercooling effect.[32] With derivation (see Section 3 in supplementary material), the heat flux $q_{rev}$ for these two thermal modules is formulated as follows.

Reverse S1:
$$q_{rev} = \frac{k_{Al} k_{Bl\_eff} (T_h - T_c)}{k_{Al} L_B + k_{Bl\_eff} L_A} \qquad (2)$$

Reverse S2:
$$q_{rev} = k_{Bl\_eff} \frac{k_{Al}(T_h - T_{mA}) + k_{As}(T_{mA} - T_c)}{k_{As} L_B + k_{Bl\_eff} L_A} \qquad (3)$$

where $k_{As}$ and $k_{Al}$ are the thermal conductivities of PCM A in solid and liquid phases, respectively; $k_{Bl\_eff}$ is the effective thermal conductivity of liquid PCM B with natural convection; and $L_A$ and $L_B$ are the lengths of terminals A and B of the SL-PCTD, respectively. The heat flux $q$ ($q_{for}$ and $q_{rev}$) of each thermal module is further amended by considering thermal contact resistances between the interfaces of the SL-PCTD (Eq. (S12) in supplementary material). By using the modified heat fluxes in the forward and reverse directions, the thermal rectification ratio in each thermal module can be determined.

A series of experimental and theoretical investigations have been systematically conducted for the forward-direction and reverse-direction operations. The $\Delta T$ implemented in the experiment is 10–40 °C (the corresponding $T_c$ being 30–0 °C). Shown in Figure 1(a) and 1(b) are the variations of the heat flux $q$ ($q_{for}$ and $q_{rev}$) and thermal resistance ($R$) of the SL-PCTD in the forward and reverse directions, where the experimental and theoretical results show a good agreement.

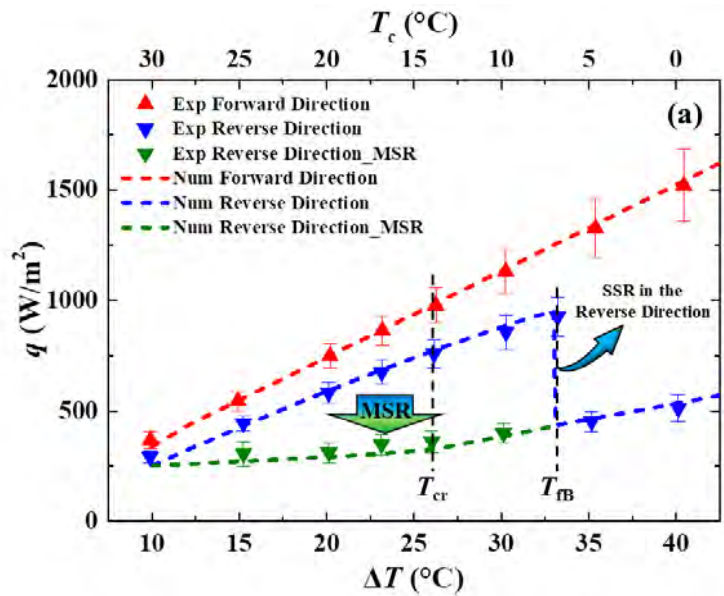

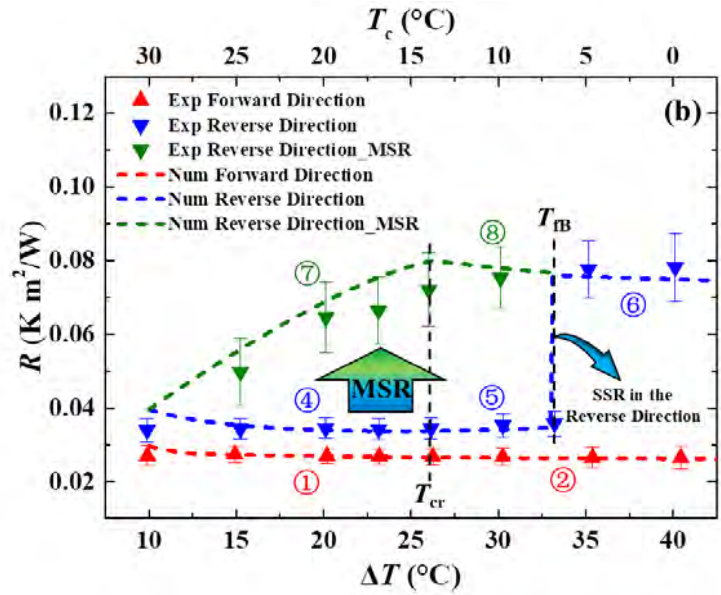

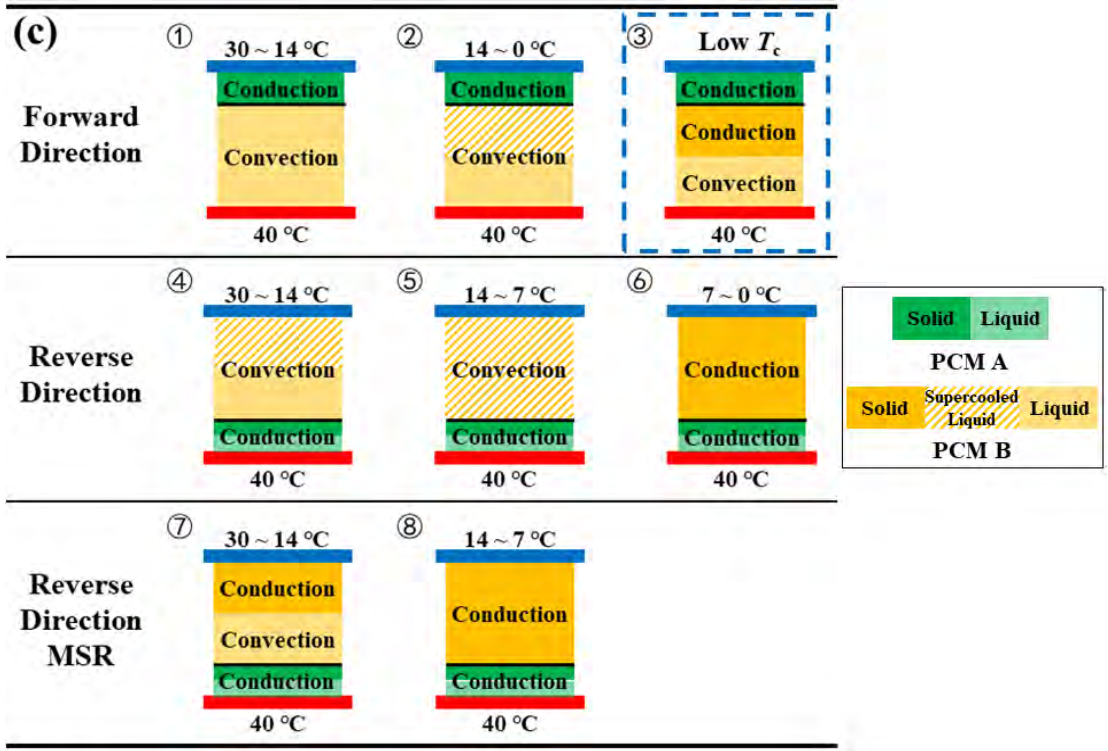

FIG. 1. Experimental (Exp) and numerical (Num) heat fluxes $q$ (a) and thermal resistances $R$ (b) of the SL-PCTD as the function of $\Delta T$ and $T_c$ in the forward and reverse directions. (c) Physical states of PCMs driven by varying $T_c$. The physical state in the dashed square is only for theoretical analysis, which was not experimentally tested due to the difficulty in steadily maintaining a very low $T_c$.

In the forward direction, as shown by red curve in Figure 1(a), the heat flux $q_{for}$ rises within the $\Delta T$ range of 10–40 °C regardless of whether $T_c$ is above or below the critical temperature ($T_{cr}$) of 14 °C–defined as the arbitrary temperature at which the interconversion of phases of PCM B is expected.[32] The increase of $q_{for}$ is attributed to small $R$ of the SL-PCTD (red curve in Figure 1(b)), which is determined by the physical states and heat transfer modes of PCMs. At $T_c > T_{cr}$ (the corresponding $\Delta T$ being 10–26 °C), PCM A retains solid phase under heat conduction, and PCM B retains liquid phase under natural convection, as depicted in Figure 1(c)–①. At $T_c < T_{cr}$ (the corresponding $\Delta T$ being 26–40 °C), PCM A still remains solid phase under heat conduction, while the upper region of PCM B is occupied by supercooled liquid phase, as presented in Figure 1(c)–②. Owing to the (supercooled) liquid phase, natural convection remains active continuously, leading to persistently small $R$ within the $\Delta T$ range of 10–40 °C. The supercooling of PCM B undergoes spontaneous supercooling release (SSR) at sufficient low $T_c$ according to theoretical analysis, where the upper region of PCM B solidifies, as shown in Figure 1(c)–③. The appearance of solid phase weakens the natural convection of PCM B, resulting in a large $R$ and small $q_{for}$. However, it is difficult to maintain a very low $T_c$ steadily in the present experiment. Therefore, only theoretical analysis of the thermal module in Figure 1(c)–③ is presented. In brief, considerable $q_{for}$ is maintained in the forward direction within the $\Delta T$ range of 10–40 °C enabled by the minimum $R$ of the SL-PCTD resulted from natural convection.

Conversely, smaller heat flux $q_{rev}$ is observed in the reverse direction, as shown by blue curve in Figure 1(a), and most importantly, two different trends are noticed when the SSR occurs at $T_c = T_{fB}$ (7 °C). The $R$ of the SL-PCTD in the reverse direction (blue curve in Figure 1(b)) is slightly larger than that of the forward direction (red curve in Figure 1(b)) at $T_c > T_{fB}$, and becomes very high at $T_c < T_{fB}$, which is based on the physical states shown in Figure 1(c)–④⑤⑥. In the entire reverse direction, PCM A always remains in the two-phase state because of $T_h > T_{mA}$, and the heat transfer in PCM A is dominated by heat conduction due to its finite length. However, PCM B undergoes various physical states. At $T_c > T_{cr}$ (the corresponding $\Delta T$ being 10–26 °C), it becomes multiphase, i.e., the region adjacent to the top cold end being supercooled liquid phase while liquid phase prevailing near PCM A, as shown in Figure 1(c)–④. At $T_{fB} < T_c < T_{cr}$ (the corresponding $\Delta T$ being 26–33 °C), a complete supercooled liquid phase of PCM B (Figure 1(c)–⑤) emerges. The heat transfer in PCM B in Figure 1(c)–④⑤ is dominated by natural convection, leading to a relatively small $R$ within the $\Delta T$ range of 10–33 °C. At $T_c = T_{fB}$ (the corresponding $\Delta T$ being 33 °C), a complete solid phase (Figure 1(c)–⑥) emerges due to SSR, resulting in maximum $R$ and minimum $q_{rev}$ in the $\Delta T$ range of 33–40 °C.

Although the heat flux $q_{rev}$ in the reverse direction (blue curve in Figure 1(a)) is relatively inhibited compared to the heat flux $q_{for}$ in the forward direction (red curve in Figure 1(a)), its rising trend is undesired particularly at $\Delta T < 33$ °C. In fact, inhibiting $q_{rev}$ as minimum as possible in the reverse direction, i.e., enhancing the $R$, is essential to efficiently improve the performance of the SL-PCTD. When $T_c$ is within 30–7 °C (the corresponding $\Delta T$ being 10–33 °C), the physical state of PCM B is (supercooled) liquid (Figure 1(c)–④⑤) wherein the natural convection enhances heat transport. By contrast, by implementing the MSR into PCM B, $q_{rev}$ in the reverse direction can be easily gated to a small value (green curve in Figure 1(a)), converting the supercooled liquid PCM B into a partial or complete solid state that enhances $R$, as shown by green curve in Figure 1(b). As a consequence, At $T_c > T_{cr}$ (the corresponding $\Delta T$ being 10–26 °C), inter-conversion of physical states of PCM B (Figure 1(c)–⑦) offers high $R$, greatly reducing $q_{rev}$. Similarly, at $T_c < T_{cr}$ (the corresponding $\Delta T$ being 26–33 °C), $q_{rev}$ tends to be low due to a complete solid phase of PCM B (Figure 1(c)–⑧). Therefore, MSR gating plays a pivotal role in inhibiting the heat flux in the reverse direction.

By using $q_{for}$ and $q_{rev}$ as discussed above, we derive the corresponding experimental and numerical thermal rectification ratios $\gamma$ (Figure 2(a)) that are defined as follows: $\gamma_{FS.RS}$ being the ratio of $q_{for}$ to $q_{rev}$, and $\gamma_{MSR}$ being the ratio of $q_{for}$ to $q_{rev}$ obtained via MSR gating. Moreover, the thermal rectification ratio ($\gamma_{FW.RW}$) of the SL-PCTD without supercooling effect from the previous work[32] is also included in Figure 2(a). The corresponding $q_{for}$ and $q_{rev}$ are given in Section 4 in supplementary material. As can be understood from Figure 2(a), $\gamma_{FS.RS}$ is just around 1.25 within the $\Delta T$ range of 10–33 °C, which is due to small $R$ on account of (supercooled) liquid phase of PCM B in the reverse direction, providing the minimum value. When the SSR of PCM B occurs in the reverse direction at $\Delta T = 33$ °C, $R$ of the reverse direction becomes large, due to which $\gamma_{FS.RS}$ around 2.9 is achieved. Using the theoretical model that has been verified by the experimental results, $\gamma$ at larger $\Delta T$ is investigated and shown in the inset of Figure 2(a). It can be found that $\gamma_{FS.RS}$ maintains around 2.8 within the $\Delta T$ range of 33–87 °C, beyond which the SSR of PCM B occurs in the forward direction, resulting in a low $\gamma_{FS.RS}$ at $\Delta T > 87$ °C. By comparing $\gamma_{MSR}$ with $\gamma_{FS.RS}$ within the $\Delta T$ range of 10–33 °C, it can be seen that $\gamma_{MSR}$ keeps on increasing within the $\Delta T$ range of 10–26 °C and eventually gains the value of 2.95 and then keeps almost constant. MSR gating introduces a great $R$ by converting supercooled liquid phase into solid phase at a higher $T_c$ within the $\Delta T$ range of 10–33 °C in the reverse direction. By combining $\gamma_{MSR}$ (green curve) within the $\Delta T$ range of 10–33 °C and $\gamma_{FS.RS}$ (blue curve) within the $\Delta T$ range of 33–100 °C, the optimum thermal rectification performance of the SL-PCTD is obtained in assistance of supercooling effect and its MSR, which shows a big increment compared to $\gamma_{FW.RW}$ (black curve) in the previous work within the $\Delta T$ range of 26–87 °C, as shown in the inset of Figure 2(a).

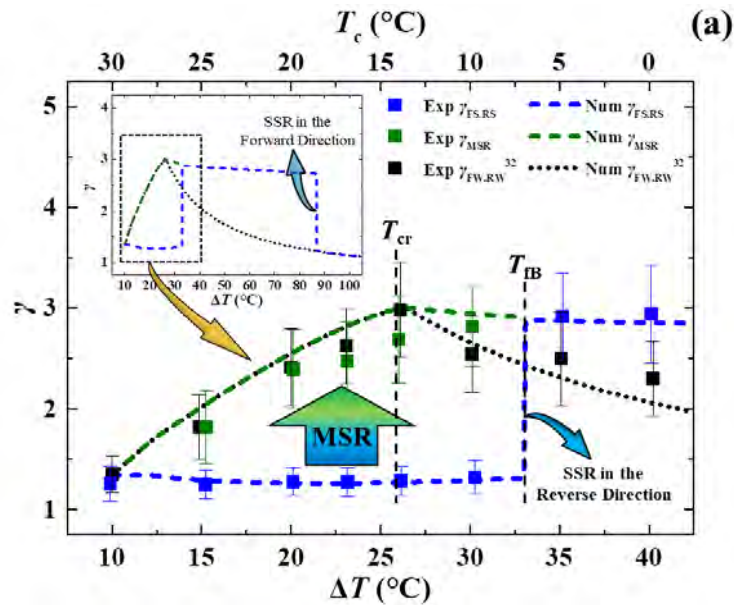

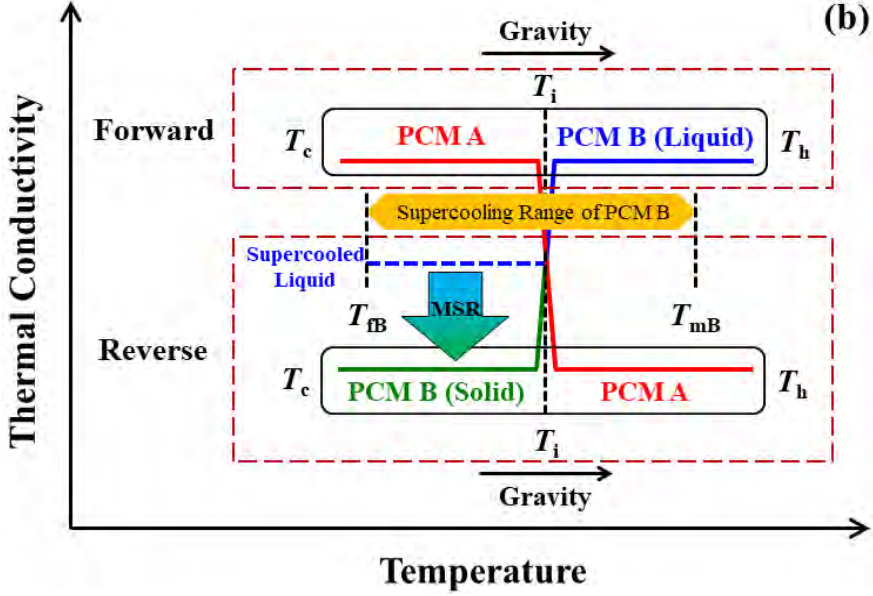

FIG. 2. (a) Experimental (Exp) and numerical (Num) thermal rectification ratios $\gamma$ as the function of $\Delta T$ and $T_c$ within the $\Delta T$ range of 10–40 °C. The numerical results within the $\Delta T$ range of 10–100 °C are shown in the inset, where the dashed square represents the $\Delta T$ range of 10–40 °C implemented in the experiment. (b) The proposed governing mechanism of a SL-PCTD under the effect of supercooling. The interfacial temperature $T_i$ of the SL-PCTD is in the temperature range of $T_{fB}$–$T_{mB}$, ensuring the supercooling of PCM B in both the forward and reverse directions. The thermal conductivities of PCM A, (supercooled) liquid PCM B and solid PCM B are schematically shown by red, blue and green lines, respectively.

The present study aims to utilize the unexplored thermo-physical properties of PCMs, i.e., supercooling and its release to enable a SL-PCTD with sustainable thermal rectification. We thus propose the governing mechanism (shown schematically in Figure 2(b)) as follows. When the interfacial temperature ($T_i$) between the two terminals is in the temperature range of $T_{fB}$–$T_{mB}$, supercooling of PCM B affects the performance of the SL-PCTD in both the forward and reverse directions. In the forward direction, PCM A works in the temperature range of $T_c$–$T_i$ with high thermal conductivity (red line within $T_c$–$T_i$). On the other hand, PCM B works in the temperature range of $T_i$–$T_h$, where PCM B retains (supercooled) liquid due to the effect of supercooling, resulting in high thermal conductivity with the assistance of natural convection (blue solid line within $T_i$–$T_h$).

Conversely, PCM A works with low thermal conductivity in the reverse direction (red line within $T_i$–$T_h$). Meanwhile, supercooling makes PCM B remain in supercooled liquid state in the temperature range of $T_{fB}$–$T_i$ with relatively high thermal conductivity (blue dashed line within $T_{fB}$–$T_i$), which is undesired for entailing a large thermal rectification ratio. However, the supercooling of PCM B can be dislodged by implementing MSR, leading to small thermal conductivity (green line within $T_{fB}$–$T_i$) and low heat flux in the reverse direction. Therefore, the thermal rectification ratio of the SL-PCTD with supercooling and MSR gating is high and sustainable compared with the conventional SL-PCTDs without supercooling.

In summary, a SL-PCTD under the effect of supercooling is presented. The findings reveal that considerably high heat flux in the forward direction is driven by intense natural convection in the liquid state, while heat conduction prevails in the solid state (high thermal resistance) in the reverse direction particularly under MSR that renders the SL-PCTD with effective gating. Consequently, thermal rectification ratio of 2.95 is obtained with sustainable trend even at large temperature biases. In all, the convincing experimental and theoretical evidences help establish the governing mechanism that combines the asymmetric thermal transport with supercooling effect and MSR.

See the supplementary material for more details about the material characterization, experimental methodology, theoretical modeling and difference in working mechanisms of the SL-PCTDs with and without the effect of supercooling.

This research is supported by the National Key Research and Development Program of China under the Contract No. 2018YFA0702300 and by the National Natural Science Foundation of China under the contract No. 51676122. A few characterizations are conducted in the AEMD of Shanghai Jiao Tong University. We thank the anonymous reviewers for the valuable comments and suggestions that improve this manuscript.

The data that support the findings of this study are available from the corresponding author upon reasonable request.

Supplementary Material

**Thermal Rectification of Solid-liquid Phase Change Thermal Diode under the Effect of Supercooling**

Zhaonan Meng, Raza Gulfam, Peng Zhang*, Fei Ma

Institute of Refrigeration and Cryogenics, Shanghai Jiao Tong University, No. 800 Dongchuan Road, Shanghai 200240, China

E-mail: zhangp@sjtu.edu.cn

## 1. Determination of freezing temperature of CaCl$_2$ solution

The freezing temperature ($T_{fB}$) of CaCl$_2$ solution was experimentally measured. CaCl$_2$·6H$_2$O (about 30 ml) was put in a beaker and was subsequently heated in a water bath at 50 °C. When the CaCl$_2$·6H$_2$O was completely melted into aqueous CaCl$_2$ solution, the beaker was put into a glycol bath maintained at -10 °C. A resistive temperature sensor was immersed in aqueous CaCl$_2$ solution to measure the $T_{fB}$ which was determined to be about 7 °C, as shown in Figure S1.

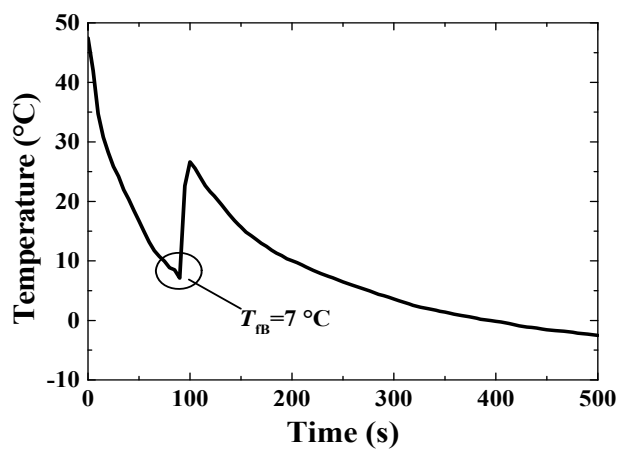

FIG. S1. Temporal evolution of temperature of aqueous CaCl$_2$ solution in cold glycol bath.

## 2. Fabrication of the SL-PCTD and experimental methodology

Two cylindrical containers made of polymethyl methacrylate were employed to fabricate two terminals of the SL-PCTD, as schematically depicted in Figure S2(a) and S2(b). Paraffin (PCM A) and $CaCl_2·6H_2O$ (PCM B) were put in the sealed chambers (labeled as terminal A and terminal B), which were then joined together via thermal conductive grease to configure the SL-PCTD. The wall thickness and outer diameter of the terminal were 2.0 mm and 50 mm, while the lengths of terminal A and terminal B were 4.8 mm and 40 mm, respectively. The top and bottom surfaces of the terminals were closed by clamping 1.0 mm thick aluminum slices by thermal adhesive. There are two compensation cavities on the lateral wall of the two cylindrical containers, which help mitigate the risks associated with thermal contraction and thermal expansion of PCMs. In addition, compensation cavity of terminal B is also used to seed solid crystals of $CaCl_2·6H_2O$ for manual supercooling release (MSR).

To configure the SL-PCTD working in the forward and reverse directions, both terminals can be interchanged, demonstrating ease of operation as well as design scalability, as shown in Figure S2(c). For the forward-direction operation, terminal A is positioned at the top and terminal B is placed at the bottom; and for the reverse-direction operation, both terminals are swapped. Under either condition, the temperature ($T_h$) of the bottom hot end of the SL-PCTD is kept constant at 40 °C, while the temperature ($T_c$) of the top cold end of the SL-PCTD is decreased from 30 °C to 0 °C, inducing the supercooling effect of $CaCl_2·6H_2O$ to emerge naturally during the cooling process in the experiments. To dislodge the supercooling effect of PCM B in the reverse-direction operation (i.e., inhibiting the heat flux), MSR is adopted in the $T_c$ range of 30–7 °C, which includes seeding solid crystals of $CaCl_2·6H_2O$ (crystal-seeding technique) into terminal B from compensation cavity, as shown in Figure S2(d). This operation results in converting supercooled PCM B into solid phase.[1]

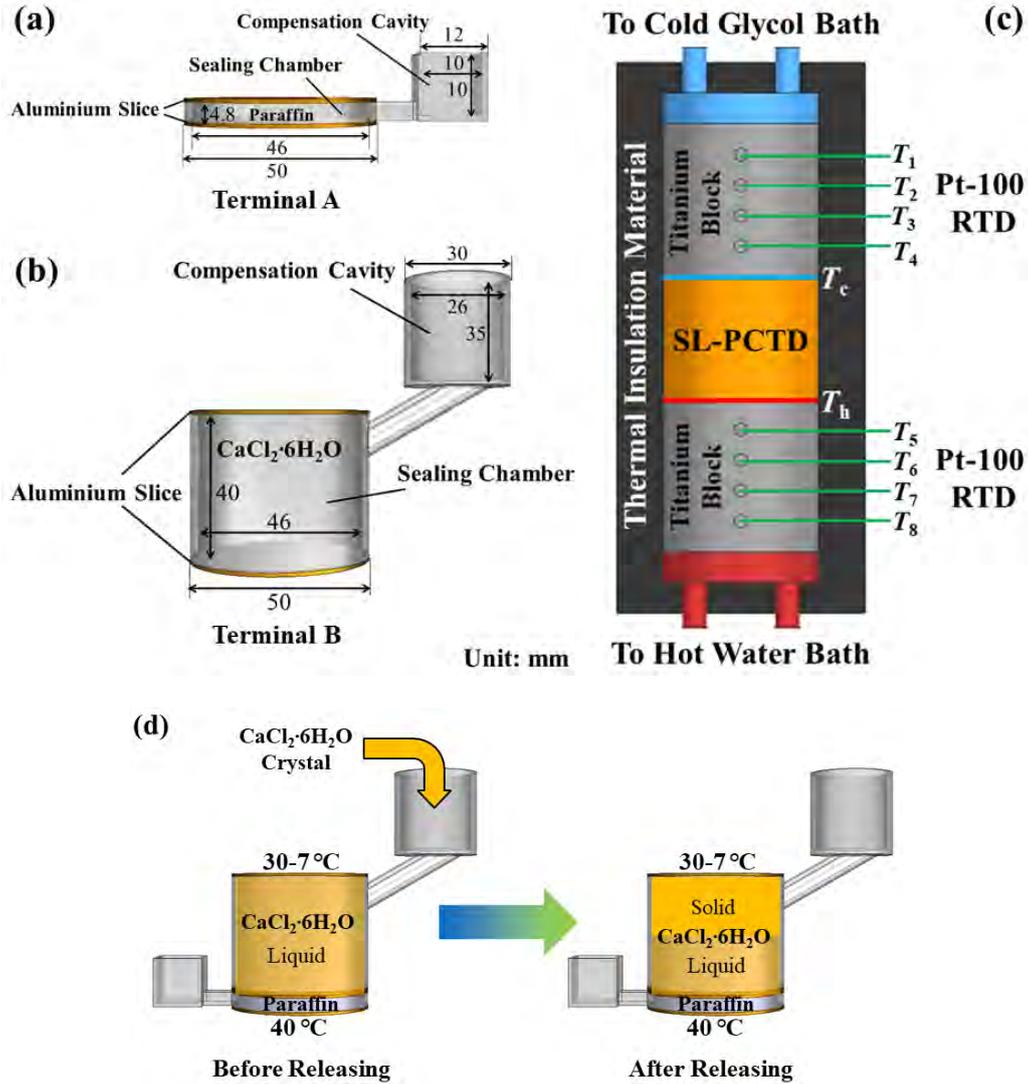

FIG. S2. Fabrication of the SL-PCTD. (a) terminal A. (b) terminal B (The dimensions are shown in the unit of mm). (c) The custom-built experimental setup to study the thermal rectification of the SL-PCTD. (d) MSR by seeding $CaCl_2·6H_2O$ solid crystals into terminal B.

In the experiments, the SL-PCTD was placed in the test section of a custom-built device,[2] as shown in Figure S2(c). The test section is clamped with two outward-extended titanium blocks with a diameter of 50 mm which are further attached to the constant-temperature bottom hot water bath and top cold glycol bath, respectively. To avoid heat loss in the experiments, the experimental devices including the SL-PCTD and titanium blocks were covered by thermal insulation materials. To measure the heat flux across the SL-PCTD, four equidistant resistive temperature sensors (RTD Pt-100) were centrally inserted into the top and bottom titanium blocks, respectively, and the distance between two adjacent RTDs is about 10.0 mm. After the steady state of the SL-PCTD was reached, the temperature history was recorded using data acquisition system (Keithley 2700), and the heat flux $q$ was calculated through the temperature gradient $dT/dx$ by means of Eq. (S1):

$$q = -k_{Ti} \frac{dT}{dx} \tag{S1}$$

where $k_{Ti}$ is the temperature-dependent thermal conductivity of titanium blocks. Using the custom-built experimental device,[3] the thermal conductivities $k_{Ti}$ at different temperatures were obtained through the steady-state method, which can be linearly fitted as follows:

$$k_{Ti} = 0.02467 T_{Ti} + 5.64955 \tag{S2}$$

where $T_{Ti}$ is the mean temperature of the titanium block.

## 3. Theoretical consideration

### 3.1. Theoretical modeling

The performance of the SL-PCTD is evaluated in terms of thermal rectification ratio ($\gamma$), as given underneath:

$$\gamma = \frac{q_{for}}{q_{rev}} \tag{S3}$$

where $q_{for}$ and $q_{rev}$ are the heat fluxes of the SL-PCTD in the forward and reverse directions, respectively.

The heat flux $q$ ($q_{for}$ and $q_{rev}$) through the SL-PCTD is assumed to be unidirectional with the exception of heat losses from the boundaries. Therefore, $q_{for}$ and $q_{rev}$ can be estimated by thermal resistance model. The SL-PCTD consists of paraffin (PCM A) and CaCl$_2$·6H$_2$O (PCM B). At different $T_h$ and $T_c$, PCM A and PCM B can exist in solid or liquid phases. Therefore, the solid phase and liquid phase of two PCMs can be treated as thermal resistances, respectively, leading to the definitions of the thermal resistances ($R$) of solid PCM A ($R_{As}$), liquid PCM A ($R_{Al}$), solid PCM B ($R_{Bs}$) and liquid PCM B ($R_{Bl\_eff}$) as follows:

$$R_{As} = \frac{L_A - \delta_A}{k_{As}} \tag{S4}$$

$$R_{Al} = \frac{\delta_A}{k_{Al}} \tag{S5}$$

$$R_{Bs} = \frac{L_B - \delta_B}{k_{Bs}} \tag{S6}$$

$$R_{Bl\_eff} = \frac{\delta_B}{k_{Bl\_eff}} \tag{S7}$$

where $L_A$ and $L_B$ are the lengths of terminal A and terminal B of the SL-PCTD, $\delta_A$ and $\delta_B$ are the heights of the PCM A and PCM B in liquid phase, $k_{As}$ and $k_{Al}$ are the thermal conductivities of PCM A in solid and liquid phases, $k_{Bs}$ is the thermal conductivity of PCM B in solid phase, and $k_{Bl\_eff}$ is the effective thermal conductivity of liquid PCM B with natural convection.

In this way, the theoretical model can be depicted with the thermal resistances in series with uniform heat flux, which is formulated as Eq. (1) in the manuscript.

### 3.2. Classification of thermal modules

According to the operation conditions of the SL-PCTD, the theoretical model can be categorized into multiple thermal modules by relating the interfacial temperature ($T_i$), $T_c$, $T_{fB}$, the phase change temperatures ($T_{mA}$ and $T_{mB}$) of PCM A and PCM B. As shown in the previous work,[2] there are four thermal modules in the forward direction and four in the reverse direction. Considering the supercooling effect of PCM B, there are two additional thermal modules (Reverse S1 and S2) in the reverse direction, as shown in Table S1.

Table S1. Schematic illustration of the two additional thermal modules in the reverse direction under the effect of supercooling. The dark blue and light blue exhibit PCM A in solid state and liquid state with thermal resistances of $R_{As}$ and $R_{Al}$, respectively; while light red shows PCM B in (supercooled) liquid state with thermal resistance of $R_{Bl\_eff}$.

| Operation conditions | $T_i \geq T_{mA}$ and $T_c > T_{fB}$ | $T_i < T_{mA}$ and $T_c > T_{fB}$ |
|---|---|---|
| Thermal modules | Reverse S1 | Reverse S2 |

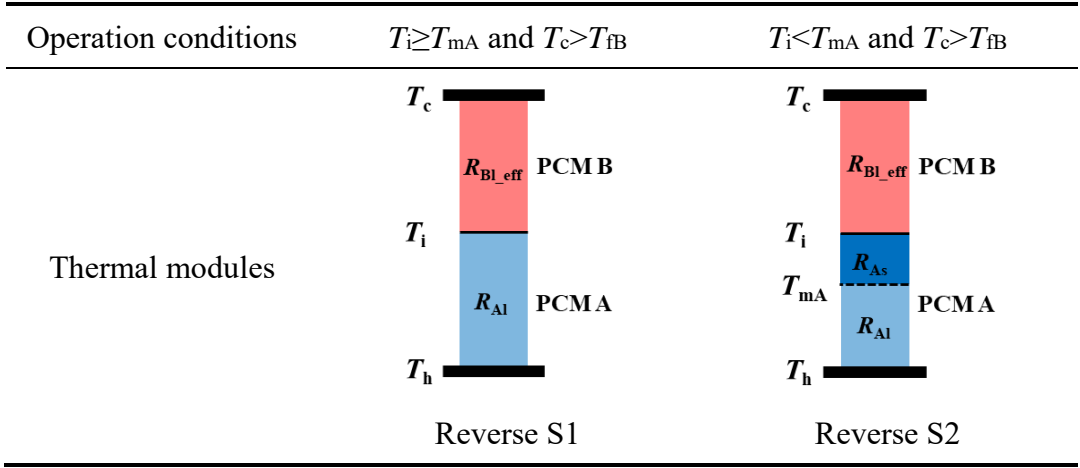

## 3.3. Solutions of heat fluxes for different thermal modules

In the theoretical model, the thermo-physical properties of PCM A and PCM B ($k_{As}$, $k_{Al}$, $T_{mA}$, $k_{Bs}$, $k_{Bl\_eff}$ and $T_{mB}$) have been determined in the previous work,[2] and the lengths of the SL-PCTD ($L_A$ and $L_B$), $T_h$ and $T_c$ are given. $T_i$, $q$ ($q_{for}$ or $q_{rev}$), the liquid heights of PCMs A and B ($\delta_A$ and $\delta_B$) for different thermal modules can be derived. The solutions of thermal modules Forward 1–4 and Reverse 1–4 have been presented in the previous work,[2] and the derivation of the two additional thermal modules (Reverse S1 and S2) are listed as follows.

In the thermal module Reverse S1, PCMs A and B only exist in liquid state, thus $\delta_A = L_A$ and $\delta_B = L_B$. Combining Eqs. (S4)–(S7) with Eq. (1) in the manuscript, we derive the governing equations of Reverse S1 as follows:

$$\begin{cases} q_{rev} = \dfrac{T_i - T_c}{\delta_B} k_{Bl\_eff} = \dfrac{T_h - T_i}{\delta_A} k_{Al} \\ \delta_A = L_A \\ \delta_B = L_B \end{cases} \quad (S8)$$

Solving Eq. (S8), we arrive at $T_i$ and $q_{rev}$ as follows:

$$\begin{cases} T_i = \dfrac{k_{Al} L_B T_h + k_{Bl\_eff} L_A T_c}{k_{Al} L_B + k_{Bl\_eff} L_A} \\ q_{rev} = \dfrac{k_{Al} k_{Bl\_eff} (T_h - T_c)}{k_{Al} L_B + k_{Bl\_eff} L_A} \end{cases} \quad (S9)$$

In the thermal module Reverse S2, PCM A exists in partial liquid and partial solid states, while PCM B exists in liquid state ($\delta_B = L_B$). Combining Eqs. (S4)–(S7) with

Eq. (1) in the manuscript, the governing equations of Reverse S2 are shown as follows:

$$\begin{cases} q_{rev} = \dfrac{T_i - T_c}{\delta_B} k_{Bl\_eff} = \dfrac{T_{mA} - T_i}{L_A - \delta_A} k_{As} = \dfrac{T_h - T_{mA}}{\delta_A} k_{Al} \\ \delta_B = L_B \end{cases} \quad (S10)$$

Solving Eq. S(10), $T_i$, $\delta_A$ and $q_{rev}$ are derived as follows:

$$\begin{cases} T_i = \dfrac{k_{Al} L_B (T_h - T_{mA}) - k_{Bl\_eff} L_A (T_{mA} - T_c)}{k_{As} L_B + k_{Bl\_eff} L_A} + T_{mA} \\ \delta_A = \dfrac{k_{Al}}{k_{Bl\_eff}} \dfrac{(T_h - T_{mA})(k_{As} L_B + k_{Bl\_eff} L_A)}{k_{Al}(T_h - T_{mA}) + k_{As}(T_{mA} - T_c)} \\ q_{rev} = k_{Bl\_eff} \dfrac{k_{Al}(T_h - T_{mA}) + k_{As}(T_{mA} - T_c)}{k_{As} L_B + k_{Bl\_eff} L_A} \end{cases} \quad (S11)$$

In order to simplify theoretical analysis, thermal contact resistances between each interface in the SL-PCTD are used to amend the above-calculated heat flux $q$ ($q_{for}$ and $q_{rev}$), resulting in heat flux $q_c$ ($q_{c\_for}$ and $q_{c\_rev}$) as given below:[2]

$$q_c = q \dfrac{R_{td}}{R_{td} + R_{c\_td}} \quad (S12)$$

where $q$ is the heat flux without the influence of thermal contact resistance, $R_{td}$ is the thermal resistance of the SL-PCTD being estimated by $R_{td}=(T_h-T_c)/q$, and $R_{c\_td}$ is the thermal contact resistance of the SL-PCTD that is equal to 0.0045 K m² W⁻¹.[2] Using the formulated $q_c$ in the forward and reverse directions, thermal rectification ratio in each thermal module can be determined.

## 4. Difference in working mechanisms of the SL-PCTDs with and without supercooling effect

The heat fluxes $q_{for}$ and $q_{rev}$ in the SL-PCTDs with (the present work) and without (the previous work) the effect of supercooling are shown in Figure S3(a) and S3(b) for comparison. As shown in Figure S3(a), $q_{for}$ in the present study (red curve) and in the previous study (black curve) are the same when $T_c$ is above the critical temperature ($T_{cr}$) of 14 °C (the corresponding temperature bias ($\Delta T$) being 10-26°C), and show different trends at $T_c < T_{cr}$. In the previous study, at $T_c < T_{cr}$ (the corresponding $\Delta T$ being larger than 26 °C), the temperature of upper region of $CaCl_2·6H_2O$ is lower than $T_{mB}$ (30 °C), and this part of $CaCl_2 \cdot 6H_2O$ is in solid state, impeding the increase of $q_{for}$ in the forward direction. In the present study, $CaCl_2$ solution still retains liquid phase owing to the supercooling, and natural convection remains active continuously, leading to the rapid increase of $q_{for}$. As shown by the numerical results in the inset of Figure S3(a), the spontaneous supercooling release (SSR) occurs in PCM B at $\Delta T$ = 87 °C, leading to a sudden drop of $q_{for}$, and then $q_{for}$ of the present work equals that of the previous work within the $\Delta T$ range of 87-100 °C.

The $q_{rev}$ in the reverse direction are shown in Figure S3(b). Comparing $q_{rev}$ in the present study (blue curve) and in the previous study (black curve), it can be found that $q_{rev}$ in the present study is larger than that in the previous study at $T_c > T_{fB}$ (the corresponding $\Delta T$ being 10–33 °C). This is attributed to the supercooling effect inhibiting the solidification of $CaCl_2$ solution, leading to a larger $q_{rev}$ than that without supercooling. However, the rising trend of $q_{rev}$ in the reverse direction is undesired for large thermal rectification ratio. Thus, MSR is implemented to reduce $q_{rev}$ in the reverse direction at $\Delta T < 33$ °C, which makes the $CaCl_2$ solution turn into a partial or complete solid state, resulting in a drastic decrease of $q_{rev}$ in the reverse direction (green curve) and the same as that in the previous study (black curve).

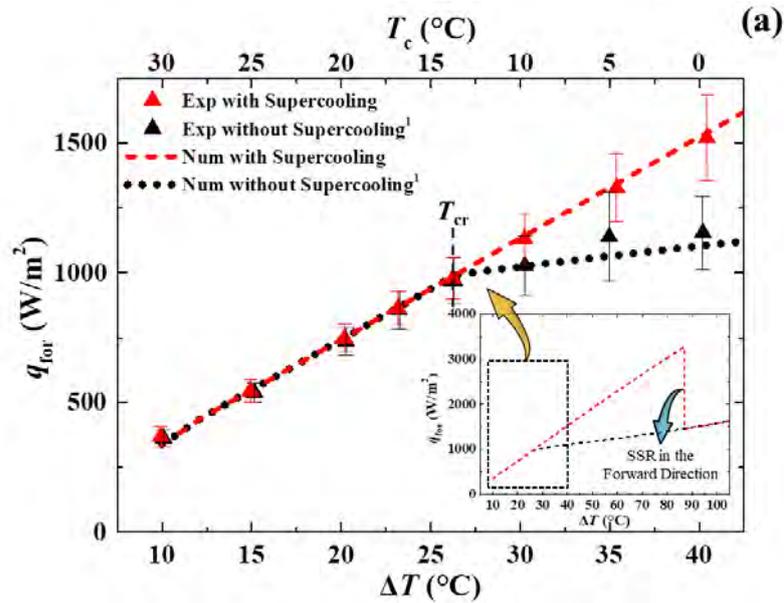
8

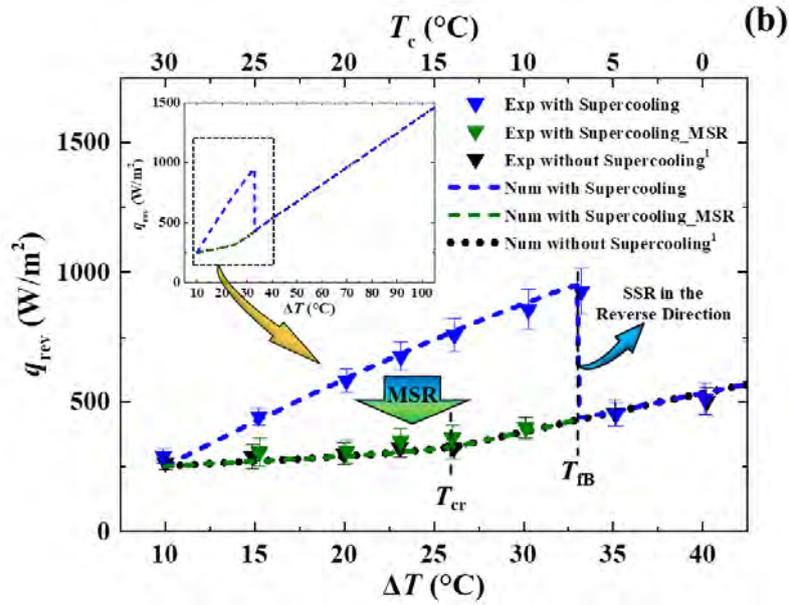

FIG. S3. (a) Experimental (Exp) and numerical (Num) $q_{for}$ of the SL-PCTD as the function of $\Delta T$ and $T_c$ within the $\Delta T$ range of 10–40 °C obtained in the present study (red curve) and in the previous study (black curve). The numerical results within the $\Delta T$ range of 10–100 °C are shown in the inset, where the dashed square represents the $\Delta T$ range of 10–40 °C implemented in the experiment. (b) Experimental (Exp) and numerical (Num) $q_{rev}$ of the SL-PCTD as the function of $\Delta T$ and $T_c$ within the $\Delta T$ range of 10–40 °C obtained in the present study (blue curve for the case under the effect of supercooling and green curve for the case with MSR) and in the previous study (black curve for the case without the effect of supercooling). The numerical results within the $\Delta T$ range of 10–100 °C are shown in the inset, where the dashed square represents the $\Delta T$ range of 10–40 °C implemented in the experiment.